\documentclass[preprint,12pt]{elsarticle}

\usepackage{amssymb}

\newcommand{\lf}{\left}
\newcommand{\rg}{\right}
\newcommand{\be}{\begin{equation}}
\newcommand{\ee}{\end{equation}}
\newcommand{\bea}{\begin{eqnarray}}
\newcommand{\eea}{\end{eqnarray}}
\newcommand{\nn}{\nonumber}
\newcommand{\ba}{\begin{array}}
\newcommand{\ea}{\end{array}}

\renewcommand{\a}{\alpha}

\renewcommand{\d}{\delta}
\newcommand{\ve}{\varepsilon}

\renewcommand{\o}{\omega}

\newcommand{\s}{\sigma}
 \newcommand{\la}{\langle}
 \newcommand{\ra}{\rangle}
 
 \newcommand{\bk}{{\bf k}}
\newcommand{\bq}{{\bf q}}

\def\bv v{{\bf v}}

\def\bp{{\bf p}}

\journal{Journal of Physics and Chemistry of Solids}

\begin{document}

\begin{frontmatter}



\title{Two-dimensional Rashba metals: unconventional low-temperature transport
properties}


\author{Valentina Brosco}
\address{IOM-CNR and Scuola Internazionale di Studi Avanzati (SISSA), Via Bonomea, 265 Trieste, Italy}
\author{Claudio Grimaldi}
\address{
Laboratory of Physics of Complex Matter, Ecole Polytechnique F\'ed\'erale de
Lausanne, Station 3, CH-1015 Lausanne, Switzerland}
\author{Emmanuele Cappelluti}
\address{ISC-CNR and Department of Physics, Sapienza University of Rome, P.le A. Moro 2, 00185 Rome, Italy}
\author{Lara Benfatto}
\address{ISC-CNR and Department of Physics, Sapienza University of Rome, P.le A. Moro 2, 00185 Rome, Italy}


\begin{abstract}
Rashba spin-orbit coupling emerges in materials lacking of structural inversion symmetry, such as heterostructures, quantum wells, surface alloys and polar materials, just to mention few examples. It yields a coupling between the spin and momentum of electrons formally identical to that arising from the weakly-relativistic limit of the Dirac equation. The purpose of the present work is to give an overview of the unconventional dc transport properties of two-dimensional metals with strong Rashba spin-orbit coupling, discussing in addition  the effects of thermal broadening.
%
%
\end{abstract}

\begin{keyword}
metals \sep interfaces \sep quantum wells \sep electrical conductivity \sep transport properties



\end{keyword}

\end{frontmatter}


\section{Introduction}
\label{intro}
Recently new materials have emerged with a remarkably strong Rashba \cite{rashba1960,bychkov1984,WinklerSpinOrbitCoupling} spin-orbit coupling. Examples range from bulk polar semiconductors, such as BiTeI,  to flat Dirac materials and surface alloys.
 These are two- or three-dimensional materials where the SO coupling can be made much larger than all other relevant energy scales, thereby reaching a regime where only carriers with a given helicity are allowed in the system, i.e. the dominant spin-orbit (DSO) coupling regime. In this regime we expect that not only spin-dependent phenomena but all the properties of the system, will be strongly affected by SO interaction, implying that they can be in principle tuned by changing the SO interaction strength, as in the celebrated Datta-Das transistor \cite{datta1990}.
In these regards, understanding whether and how spin-orbit coupling affects the charge transport properties  of materials turns out to be particularly relevant.
Besides its fundamental interest, answering this question would indeed allow to clarify if the electrostatic control of Rashba coupling may be used to manipulate the conductivity of a Rashba metal and, at the same time,  if signatures of  Rashba coupling can be found in transport measurements.

To this end, in the present work we give an overview of the unconventional dc transport properties characterizing materials in the DSO regime, first studied by us in Refs.\cite{brosco2016,brosco2017}, discussing in addition  the effects of thermal broadening.

In this paper we will focus on the low-temperature transport properties of metals determined by the elastic scattering of electrons from static impurities. 
In weakly disordered samples,  when the
wavelength of the electrons is much smaller  than their  mean free path, transport can be described  quasi-classically.  The DC conductivity, $\sigma$, then  follows Drude's law
\be
\label{sdrude}
\s=\frac{ne^2\tau_{\rm tr}}{m},
\ee
where $n$ is  the density of charge carriers, $m$ their effective mass and $\tau_{\rm tr}$ denotes the transport scattering time.
%
%
%
For an ordinary two-dimensional metal, at low temperatures, when all inelastic processes are thermally suppressed, $\tau_{\rm tr}$ is a constant, {\sl i.e.} $\tau_{\rm tr}=\tau_0$, and the  conductivity scales linearly with the electronic density. 

In two recent works \cite{brosco2016,brosco2017}, focused respectively on two- and three- dimensional systems,  we showed that in the presence of Rashba spin-orbit coupling this picture breaks down and $\sigma_{\rm dc}$ has a non-trivial dependence on the Rashba energy and  on the electronic density even at $T=0$.  More specifically, we demonstrated that, in the  DSO regime, when the energy associated with Rashba coupling,  $E_0$, overcomes the Fermi energy, or equivalently the electron density goes below a threshold value $n_0=2m E_0/(\pi \hbar)^2$, $\sigma_{\rm dc}$ obeys the approximate formula in two dimensions:
\bea
\sigma_{\rm DSO }
&\simeq&
\frac{e^2 \tau_0}{m}\frac{n^2}{2 n_0}\lf(1+\frac{n^2}{n_0^2}\rg)
\quad n\leq n_0,
\label{sigmasso}
\eea
while at $n>n_0$ the dc conductivity $\sigma_{\rm dc}$ coincides with $\sigma^0_{\rm Drude}\equiv ne^2\tau_0/m$, i.e. Eq.(\ref{sdrude}) when $\tau_{\rm tr}=\tau_0$.

The non-linearity of
$\sigma$ as a function of $n$ provides a clear signaturefor the
DSO regime that can be relevant for a wide class of two-dimensional metals including surface alloys \cite{ast2007,ast2008,gierz2009, mirhosseini2010,yaji2010,gruznev2014,sanchez2013}, layered bismuth tellurohalides \cite{eremeev2012,bahramy2012,sakano2013,xi2013,chen2013},  HgTe quantum wells \cite{gui2004} and interfaces  between complex oxides \cite{ohtomo2004,thiel2006,reyren2007,caviglia2008,bell2009,biscaras2012,seri2012,benshalom2010,caviglia2010,joshua2013,liu2013,biscaras2014,zhong2015,zhong2013,joshua2012}. Indeed, in all these cases the lack of inversion symmetry induces a Rashba SO coupling \cite{WinklerSpinOrbitCoupling,bychkov1984} considerably stronger than what found in traditional III-V
 semiconductor heterostructures \cite{sulpizio2014,rashba2012}. In addition, the 
 carrier density can be tuned by means of gate field effect or chemical doping, allowing one to use the predicted behavior  (\ref{sigmasso}) of $\sigma_{\rm dc}$ as a function of $n$ to estimate the Rashba energy $E_0$.
 As we shall see, Eq.(\ref{sigmasso}) stems from the appearance\cite{cappelluti2007} at $n<n_0$
of an 
``anomalous'' contribution to the  current, that  instead vanishes, due to vertex corrections when $n>n_0$.

%
%
%

\section{Model}
\label{model}
We consider a two-dimensional electron gas (2DEG) in the presence of disorder and Rashba spin-orbit coupling. The Hamiltonian can be written as follows:
\be
H=\sum_{{\bf k}}c^\dag_{{\bf k}}h_{\bf k}c_{{\bf k}}+H_{\rm imp},
\ee
where $c^\dag_{{\bf k}}=(c^\dag_{\bk\uparrow}, c^\dag_{\bk\downarrow})$ and $c_{{\bf k}}=(c_{\bk\uparrow}, c_{\bk\downarrow})$ are spinor creation and annihilation operators, $h_{\bk}$ denotes the Hamiltonian of the clean Rashba model and  $H_{\rm imp}$ accounts for the effects of disorder.
Indicating as $\alpha$ the intensity of the Rashba coupling and $m$ the effective electron mass, $h_{\bf k}$ can be written as
\be
h_{\bf k}=\frac{k^2}{2m}\sigma_0+ \alpha\lf[{\bf k}\times\vec \sigma\rg]_z\ee
where $\vec \sigma$ is the vector of Pauli matrices and $\sigma_0$ is the $2\times2$ identity matrix.
Assuming static non-magnetic impurities $H_{\rm imp}$ can  in turn be cast as
\be
H_{\rm imp}=\sum_{\bq,\bk} c^\dag_{\bk+\bq} V_{\rm imp}(\bq) c_\bk
\ee
with $V_{\rm imp}(\bq)$ denoting the Fourier transform of the impurity potential. 

 \begin{figure}[!t]
\begin{center}
\includegraphics[width = 0.7\textwidth]{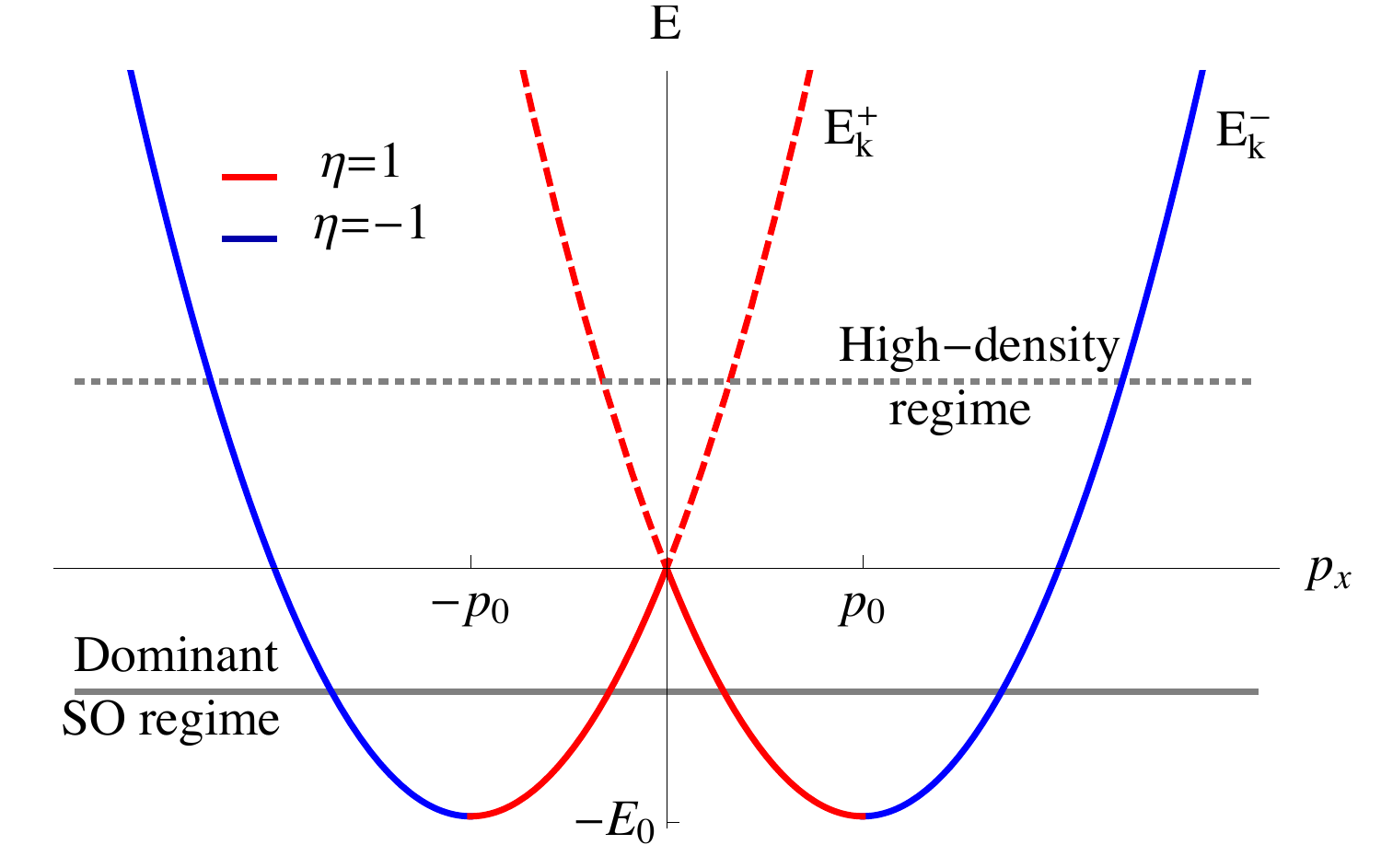}
\caption{Spectrum of the Rashba model.}\label{spectrum}
\end{center}
\end{figure}

\subsection{Quasi-particle spectrum}
The clean Rashba Hamiltonian can be diagonalized by switching to the helicity eigenstates basis.
The helicity operator is defined as $S=\lf[\hat p\times\vec \sigma\rg]_z$ and its eigenstates, $| \bk s\ra$, with $s=\pm 1$,  satisfy the  relation $S | \bk\, \pm\ra =\pm | \bk\, \pm\ra $.
A simple calculation shows  in particular that $|\bk\, \pm\ra$ can be expressed in terms of the standard spin eigenstates, $|\bk \uparrow\ra$ and $|\bk \downarrow\ra$, as 
$|\bk\, \pm\ra=(\exp{(-i \theta_{\bk})}|\bk \uparrow\ra\pm i |\bk \downarrow\ra)/\sqrt{2}$ with $\theta_\bk=\arctan(k_y/k_x)$.  Consequently,  the matrix $U_\bk$  which implements the rotation from the spin to the helicity eigenstates basis has the form
\be
U_{\bk}=\frac{1}{\sqrt{2}}
\lf(\begin{array}{cc}
e^{-i \theta_\bk} & e^{-i \theta_\bk} \\
i & -i \\
\end{array}\rg).
\ee

As stated above,  the transformation $U_{\bk}$, diagonalizes the clean Rashba Hamiltonian  and it yields an electronic spectrum consisting of two bands with dispersion 
\be E^{\pm}_k-E_0=(k\pm p_0)^2/(2m)-E_0\ee
where $E_0=m \alpha^2/2$ while $p_0=m\alpha$.
As one can see in Figure \ref{spectrum} $E_0$ corresponds to the energy difference between the lower band-edge and the ``\textit{degeneracy point}'' where the two bands touch. 
In the following we measure the Fermi energy from the lower band edge so that $E_F=E_0$ corresponds to the degeneracy point
and the threshold density $n_0$ is approximately (apart from band-renormalization effects due to disorder) the electronic density corresponding to $E_F=E_0$.
In Fig.\ref{spectrum} we can clearly distinguish two regimes, the high-density (HD) regime corresponding to  $E_F>E_0$ where both chiral bands are occupied and the dominant spin-orbit (DSO) regime, $E_F<E_0$, where only states with negative chirality are allowed at zero temperature. 
In both cases the Fermi surface consists of two circles, however,  in the HD regime states belonging to the two Fermi circles have opposite helicities while in the DSO regime they belong to the same helicity band.
As we show below and as it has been discussed in Ref.\cite{brosco2016}, this fact yields  dramatic changes in the transport and single-particle properties as the Fermi level is swept across  the degeneracy point.
To describe these changes, in a compact fashion it would be highly desirable to have a unified classification of the the eigenstates across the different regimes. To this end, following the route outlined in Ref. \cite{brosco2016},  we introduce the concept of transport helicity.
%
 \subsection{Velocity and transport helicity}
 \label{sect-tr-hel}
 The transport helicity operator, $\hat \eta$ is defined as:
 \be
 \label{defeta}
 \hat \eta=\rm{sign}(\vec v\cdot \hat k)\cdot S
 \ee
where  $S$ is the standard helicity operator and $\vec {\rm v}=\nabla_{\bf k}H$ is the velocity operator.  In the presence of Rashba spin-orbit coupling the latter acquires a spin structure and
in the spin basis it is given by 
\bea
\label{jbare}
{\rm v}_x&=&\frac{k_x}{m}\sigma_0+\alpha \sigma_y\\
{\rm v}_y&=&\frac{k_y}{m}\sigma_0-\alpha \sigma_x.
\eea%
 while in the helicity basis, it is represented  by the following matrix:
\be\label{jlongitudinal-transverse} \lf[\vec {\rm v}\rg]_{ss'}=\vec v_{{\bf k}s}\delta_{ss'}-i\alpha s(1-\delta_{ss'})\hat t_k ,
\ee%
 where $\vec v_{{\bf k}s}=\nabla_{\bf k} E^s_k=\hat k(k/m+s\alpha)$
 denotes the quasi-particle velocity and $\hat t_k$ is defined as
 $\hat t_k=\{ k_y/k, - k_x/k\}$.

\begin{figure}[b!]
\begin{center}
\includegraphics[width = 0.5\textwidth]{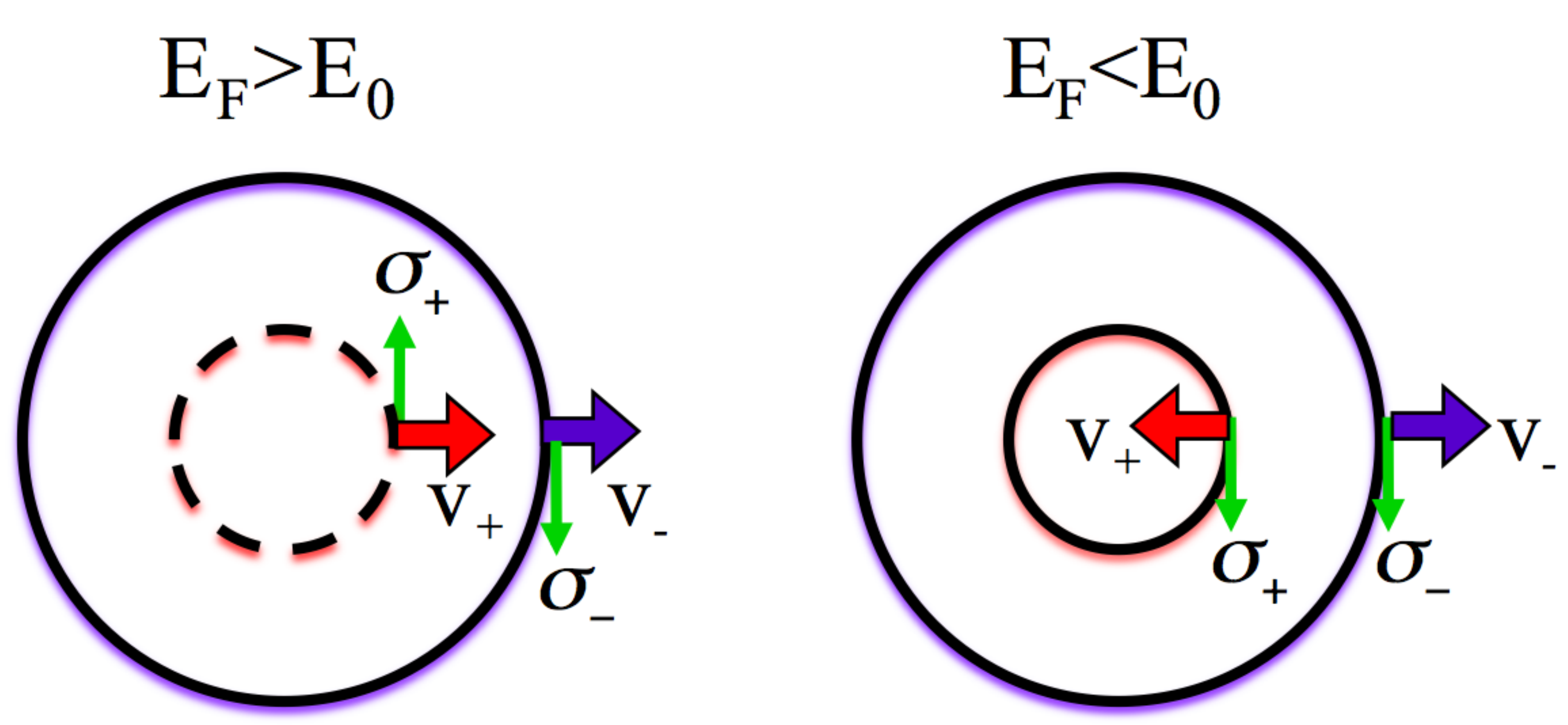}
\caption{Quasi-particle spin and velocity on the inner and outer Fermi surfaces corresponding to  transport helicity quantum numbers $\eta=1$ and $\eta=-1$, according to the definition (\ref{defeta}). }\label{velocity}
\end{center}
\end{figure}

From the above equations, 
it is straightforward to show that the operator $\hat \eta$ commutes with the Hamiltonian of the clean system
and it allows to classify univocally the states belonging the two Fermi circles both in the HD and in the DSO regime. The inner Fermi circle indeed  always corresponds to $\eta=1$ while the outer one corresponds to $\eta=-1$.
More precisely, states on the outer Fermi circles always have negative helicity and the velocity parallel to the momentum, while on the inner Fermi both the helicity and the longitudinal component of the velocity change sign as the Fermi level crosses the degeneracy point, see Fig.\ref{velocity}.
Beside the helicity eigenstates basis, $\{|{\bf k}s\ra\}$ we can thus define a transport helicity basis as $\{|E\,{\hat k}\,\eta\ra\}$.
As first shown in Ref.\cite{brosco2016}, the latter allows one to give a unified description of the transport properties of the Rashba model.
%
%

%
  \section{Born scattering in the presence of Rashba spin-orbit coupling}
Disorder effects can be accounted for both in the spin and in the helicity basis. 
In the helicity basis spanned by  the states $| \bk s\ra$,  the impurity Hamiltonian has the form
\be
H_{\rm imp}=
\sum_{\bq,\bk} V_{\rm imp}(\bq) a^\dag_{\bk+\bq} U_{\bk+\bq}^\dag U_\bk a_\bk
\ee
where  $a_\bk=(a_{\bk+},a_{\bk-})$ and $a^\dag_\bk=(a^\dag_{\bk+},a^\dag_{\bk-})$  destroy and create electrons in a given helicity state.
The above equation implies that to each impurity-scattering vertex, changing the electron momentum from ${\bf k}$ to ${\bf p}$, one has to associate the spin rotation $U^\dag_{\bf p}U_{\bk}$. In spite of this apparent complication, as shown in detail in the supplementary of Ref.\cite{brosco2016}  the choice to work in the helicity-basis allows for a simpler identification of the most relevant terms that appear in the final result and a more transparent comparison with Boltzmann approach. On the other hand the calculation in the spin basis  allows to exploit the symmetry of the model to simplify the integrals  and  it can be most easily generalized to situations where the spectrum is non-isotropic  such as in the presence of a magnetic field and in the case of bulk Rashba metals \cite{brosco2017}.
Here we start by giving an outline of the calculation in the spin basis and we use the helicity basis mainly to discuss the final results and  illustrate their physical meaning.
We assume that disorder is generated by diluted short-range impurities  such that we can set $V_{\rm imp}(\bq)=1/{\cal V}\,\sum_j e^{i \bf{q \cdot R}_j} v_{\rm imp}$  with ${\bf R}_j$ denoting the position of the impurities and 
$$\la V_{\rm imp}(\bq)V_{\rm imp}(\bq')\ra_{\rm imp}\simeq (2\pi)^2 n_{ i}v^2_{\rm imp}\delta(\bq+\bq')$$
where $\la...\ra_{\rm imp}$ denotes the average over the impurity position and $n_{i}$ is the impurity density.
Under these assumptions, within Born approximation, the effects of disorder on the system can be quantified by the elastic scattering rate in the absence of spin-orbit coupling,
$ \Gamma_0=mn_{i}v^2_{\rm imp}/2.$
\subsection{Green's function}
%
%
The Green's function obeys the standard Dyson equation:
$G^{-1}=\lf(G^0\rg)^{-1}-\Sigma$,  where  $G^0$ is the Green's function of the Rashba model in the absence of disorder and $\Sigma$ is the self-energy.

Specifically, in the spin basis, $G^0$ is represented by the following matrix: 
\be G^0=\frac{1}{2}\sum_s g_s(i\ve_l,\mathbf{k})\left[1+s(\hat{z}\times\hat{k})\rg]\label{G0}\ee
%
where $\ve_l$ is a fermionic Matsubara frequency and  $g_s(i\ve_l,\mathbf{k})=(i\ve_l-E_{ps}+E_F)^{-1}$  is the bare Green's function of each chiral eigenstate.  

\begin{figure}[t]
\begin{center}
\includegraphics[width = 1.5 cm]{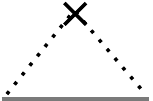}
\caption{Wigwam diagrams which describe the self-energy within Born approximation. The solid line  correspond to the dressed Green function, $G$, the dashed lines to scattering with impurities, while the crosses indicate averaging over disorder \cite{MahanManyParticlePhysics,Bruus}.}\label{wigwam}
\end{center}
\end{figure}

Within the self-consistent Born approximation (SCBA), $\Sigma({\bf p},i \ve_n)$ is determined by solving the following equation
\begin{equation}
\label{sigmascba}
\Sigma({\bf k},i \epsilon_n)_{}=\frac{n_i v_{\rm imp}^2}{\cal V}
\sum\nolimits_{{\bf p}} G({\bf p},i\epsilon_n)
\end{equation}
%
which corresponds 
the ``wigwam diagram''  depicted in Fig. \ref{wigwam} as described {\sl e.g.} in Refs.\cite{MahanManyParticlePhysics,Bruus}. %
 Note that equation (\ref{sigmascba}) is in general a matrix equation, however for our simple model, it admits a momentum- and spin- independent solution. Indeed, setting $
 \lf[\Sigma({\bf p},i \epsilon_n)\rg]_{\alpha\beta}=\Sigma(i \epsilon_n)\delta_{\alpha\beta}$, the momentum dependent part on  the r.h.s. of this equation, coming from the second term in equation (\ref{G0}), averages to zero at each order. 
%

By analytical continuation to real frequencies of Eq.(\ref{sigmascba}) (see {\sl e.g.} \cite{Bruus}) we obtain 
the following self-consistent equations for the the retarded self-energy $\Sigma^R(\omega)$ 
\be\label{sigma}
\Sigma^R(\o)=\frac{n_i v_{\rm imp}^2}{2\cal V}
\sum\nolimits_{{\bf p}s}g_s^R(p,\o) \theta(p_c-p)
\ee
and for the scattering rate $\Gamma$
\be
\label{gamma}
\Gamma=-{\rm Im}[\Sigma^R(0)]=\frac{n_i v_{\rm imp}^2 \Gamma}{2\cal V} \sum_{\bp} 
\lf[|g_+^R(p,0)|^2+|g_-^R(p,0)|^2\rg]
\ee
%
%
where $g_\pm^R(p,\omega)=\left[\o-E^\pm_p+E_F-\Sigma^R(\omega)\right]^{-1}$.
To simulate a finite Brillouin zone, in Eq. \ref{sigma}  we introduced an upper momentum cut-off, $p_c$.
The latter is needed, in particular, to regularize  the real part of the self-energy, 
${\rm Re}[\Sigma^R(\omega)]$, which would otherwise diverge logarithmically at the band edge, see {\sl e.g.} Ref.\cite{knigavko2005}.  
In these regards, we notice that, contrarily to what happens in standard half-filled systems where ${\rm Re}[\Sigma^R(\omega)]$ is approximately $\omega$-independent and it can be absorbed in a redefinition of the Fermi level, in the low-doping regime investigated here ${\rm Re}[\Sigma^R(\omega)]$ acquires a non-trivial frequency dependence. We thus need to  calculate self-consistently both ${\rm Re}[\Sigma^R(\omega)]$ and ${\rm Im}[\Sigma^R(\omega)]$.
Such self-consistent solution yields the elastic scattering rate,  $\Gamma=-{\rm Im}[\Sigma^R(0)]$ and the renormalized density of states (DOS)
$N(E)=-\frac{1}{\pi{\cal V}}\sum_{\bf p}{\rm Im} [G_R({\bf p},E)]$. The electronic density at  a temperature $T$ is given by  
\be\label{eq:nT}
n=\int_{-\infty}^{\infty} f(\ve,T ) N(\ve)  d\ve.
\ee
where $f(\ve,T)$ is the Fermi function, $f(\ve,T)=1/(e^{\beta(\ve-\mu)}+1)$ with $\beta=k_B T$ and $\mu$ indicating the chemical potential.
\subsection{DC transport within Kubo linear response theory}
\begin{figure}[t!]
\begin{center}
\includegraphics[width = 6.5 cm]{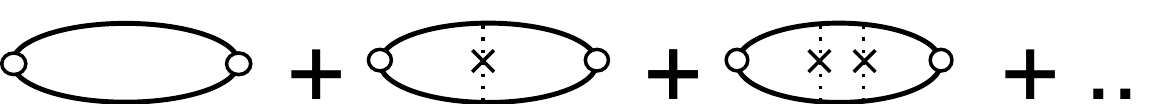}
\caption{Ladder diagrams describing the conductivity within Born approximation. Empty circles represent $j_x(\bp)$.}\label{fig-ladder}
\end{center}
\end{figure}

Now that we set up the stage for the self-consistent calculation of the Green's function, let us focus on the dc conductivity, $\s_{dc}$.
We calculate $\s_{dc}$ using Kubo linear response theory and diagrammatic perturbation theory  in Matsubara frequencies, as described in various textbooks (see {\sl e.g.} Refs. \cite{MahanManyParticlePhysics,Bruus}).

%
Within the SCBA this implies the summation of all ladder diagrams shown in Fig. \ref{fig-ladder} and it corresponds to calculate  the following 
sum 
\be\label{eq:matsusum}
-\frac{1}{\beta}\sum_{\ve_l}P_{xx}(i\ve_{l},i\ve_{l+n})
\ee
with $P_{xx}(i\ve_{l},i\ve_{l+n})$ denoting  the current-current correlation  function,
\be
\label{piba}
P_{xx}(i\ve_{l},i\ve_{l+n})=\frac{1}{{\cal V}}\sum_{{\bf p}}{\rm Tr}\left\{
 {G }({\bf p},i\ve_l )j_x({\bf p}){G }({\bf p},i\ve_{l+n})J_x({\bf p},i\ve_{l},i\ve_{l+n}) \right\}
\ee
where $\ve_{l+n}$ and $\ve_{l}$ are fermionic Matsubara frequencies, the trace is over the spin variables  and $j_x$ is the bare current, $j_x=e{\rm v}_x$. 

\begin{figure}[t]
\begin{center}
\includegraphics[width = 5 cm]{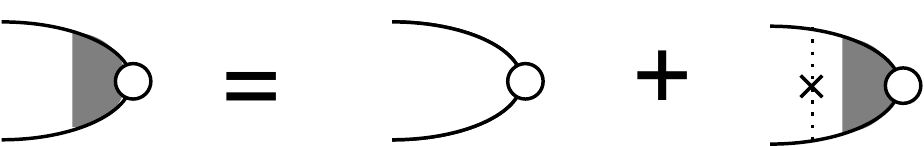}
\caption{Diagrams describing renormalization of the charge current vertex. }\label{fig-svren}
\end{center}
\end{figure}

In Eq.(\ref{piba}) the renormalized charge current,  $J_{x}({\bf p},i\ve_{l},i\ve_m)$, is  given by

\bea \label{jxkmatsu} 
J_x({\bf k},i\ve_l,i\ve_m)=j_x({\bf k})+\frac{n_i v_{\rm imp}^2}{\cal V}\sum_{\bf p}\lf[{G}({\bf p},i\ve_l)J_x({\bf p},i\ve_l,i\ve_m) {G}({\bf p},i\ve_m)\rg]
\eea
 which corresponds to the diagrammatic equation shown in Fig. \ref{fig-svren}. 
 
Starting from  Eqs.(\ref{eq:matsusum}-\ref{piba})  performing the sum over Matsubara frequencies and after analytic continuation, in the zero-frequency limit, at a temperature $T$, we obtain the following expression for the conductivity
\be\label{sigmaxx}
\tilde \sigma_{dc}=-\frac{1}{2\pi}\int  \frac{{\rm d}f(\ve,T)}{{\rm d}\ve} \lf(P_{xx}^{AR}(\ve, \ve) -{\rm Re}[P_{xx}^{RR}(\ve, \ve)]\rg) {\rm d}\ve
\ee
where the superscript $AR$ ($RR$) indicate that the first argument is advanced (retarded) and the second retarded, {\sl i.e.} 
$P_{xx}^{AR}(\ve,\ve)=P_{xx}(\ve -i\d,\ve+i\d)$ and $P_{xx}^{RR}=P_{xx}(\ve+i\d,\ve+i\d)$.

Before coming to the discussion of the results, let us rewrite the conductivity in terms of simple integrals.
The self-consistent equation for the current can be solved noting that, due to symmetry arguments the renormalized current  $J_x$ must have  the same matrix structure of the bare current \ref{jbare}, so that 
we can write:

\be
\label{jdressedmatsu}
J_x(\bk, i\epsilon_l,i\epsilon_m)=
\frac{p_x}{m}\sigma_0+\tilde\alpha( i\epsilon_l,i\epsilon_m)\sigma_y
\ee
where $\tilde \a(i\ve_l,i\ve_m)$ satisfies the following self-consistent equation:%
\be
\label{tildealpha}
\tilde\alpha= \alpha+\frac{n_iv_{\rm imp}^2}{2}\sum_{\bf p}{\rm Tr}\big[\sigma_y{G}(\mathbf{p},i\ve_l)J_x(\mathbf{p},i\ve_l,i\ve_m)\,{G}(\mathbf{p},i\ve_m)\big].\ee

Equation (\ref{tildealpha}) in turn yields the following result for the renormalized anomalous vertex:
\be\label{gcx}
\tilde\alpha(i\ve_l,i\ve_m)=\frac{\a+\a_0(i\ve_l,i\ve_m)}{1-A(i\ve_l,i\ve_m)}
\ee
where we introduced the quantities   $A(i\ve_l,i\ve_m)$ and $\a_0(i\ve_l,i\ve_m)$ given by:
\bea
\!\!\!\!\!\!\!\!\!A(i\ve_l,i\ve_m)&=&\frac{n_iv_{\rm imp}^2}{4 \cal V}\sum_{\bk s s'} {g}_s(k,i\ve_l){g}_{s'}(k,i\ve_m),\label{Aivel}\\
\!\!\!\!\!\!\!\!\!\a_0(i\ve_l,i\ve_m)&=&\frac{n_iv_{\rm imp}^2}{4\cal V}\sum_{{\bf k}s} \frac{k}{m} s\, {g}_s(k,i\ve_l){g}_s(k,i\ve_m).\label{alpha0}
\eea

By replacing Eqs.(\ref{jbare}) and (\ref{jdressedmatsu}) in Eq.(\ref{piba}),  the correlation function $P_{xx}(i\ve_{l},i\ve_m)$ can be in turn expressed as follows
\be\label{pxxivel}
P_{xx}(i\ve_l,i\ve_m)=P_0(i\ve_l,i\ve_m)+\frac{m(1-A(i\ve_l,i\ve_m))}{\Gamma_0} \tilde \alpha(i\ve_l,i\ve_m)\left(\tilde \alpha(i\ve_l,i\ve_m)-\alpha\right)
\ee
where  
we introduced the function $P_0(i\ve_l,i\ve_m)$,
\bea
\label{p0}
	P_0(i\ve_l,i\ve_m)&=&\frac{1}{2\cal V}\sum_{{\bf k}s}\frac{k^2+s k k_0}{m^2}{g}_s(k,i\ve_l){g}_s(k,i\ve_m),
\eea
that  in the absence of Rashba coupling yields the only non-vanishing contribution to the conductivity.

\section{Zero temperature results}
The equations given above allow for a fully  self-consistent numerical calculation of single-particle and transport properties of a two-dimensional electron gas with Rashba spin-orbit coupling. In the zero-temperature limit the properties and the physics of such self-consistent solution have been discussed in Ref.\cite{brosco2016}.
There we also showed that accurate, approximate analytical results for the self-energy and the conductivity can be obtained  in the weak-disorder limit (WDL), $\Gamma_0\ll E_F,E_0$. In this limit  we can in particular (i) approximate the spectral functions ${\cal A}_\pm(p,\omega)\equiv -(1/\pi) {\rm Im} g^R_\pm(p,\omega)$ with a delta, {\sl i.e.} set ${\cal A}_\pm(p,\omega)=(\Gamma/\pi)|g_+^R(p,\omega)|^2=\delta(\omega-E_\bp^\pm+E_F)$; (ii) neglect the RR contributions in Eq.(\ref{sigmaxx}).

\noindent The purpose of this Section is to give an overview of the numerical and analytical results in the zero temperature limit.
A detailed derivation of the analytical results may be found in the appendix of Ref.\cite{brosco2016} and in Ref. \cite{brosco2017}.
Here we limit ourselves to  schematically summarize the analytical results in Table \ref{table2}.

%

%
{\center
{\setlength\arraycolsep{40pt}
\begin{table}[t]
\begin{tabular}{|c|c|c|}
\hline
\rule{0pt}{5ex} 
  &   \begin{tabular}{c}
DSO regime\\
$E_F<E_0$, $n<n_0$ 
\end{tabular}&  \begin{tabular}{c}
HD regime\\
$E_F>E_0$, $n>n_0$ 
\end{tabular}\\[0.1cm]
\hline
\rule{0pt}{5ex} 
DOS, $N(E)$ & $N_0\sqrt{E_0/E}$ & $N_0$\\[0.2cm]
 \hline
\rule{0pt}{5ex} 
 \begin{tabular}{c}
Quasiparticle\\
scattering rate, $\Gamma$
\end{tabular}  & $\Gamma_0\sqrt{E_0/E_F}$ & $\Gamma_0$\\[0.2cm]
\hline
\rule{0pt}{5ex} 
Density, $n$ & $2N_0 \sqrt{E_F E_0} $&  $N_0(E_F+E_0)$\\[0.1cm]
\hline
\rule{0pt}{5ex} 
 \begin{tabular}{c}
Anomalous\\
vetex $\tilde \alpha^{RA}$
\end{tabular} &  \begin{tabular}{c}
$\alpha(1-E_F/E_0)$,\\[0.1cm]
 {\sl i.e.} $\alpha(1-n^2/n_0^2)$
\end{tabular}  & 0\\[0.1cm]
\hline
\rule{0pt}{5ex} 
 \begin{tabular}{c}
Dc conductivity\\ $\sigma_{\rm dc}$
 \end{tabular} & $\sigma_{n_0}/2\lf[(n/n_0)^4+(n/n_0)^2\rg]$ & $\sigma_{\rm Drude}$\\[0.1cm]
\hline
\end{tabular}
\caption{Analytical results in the weak disorder limit, $N_0=m/\pi$ and $\sigma_{n_0}=n_0\tau_0/m$}
\label{table2}
\end{table}}
}

%

 \begin{figure}[b!]
\begin{flushleft}
\includegraphics[width=5cm]{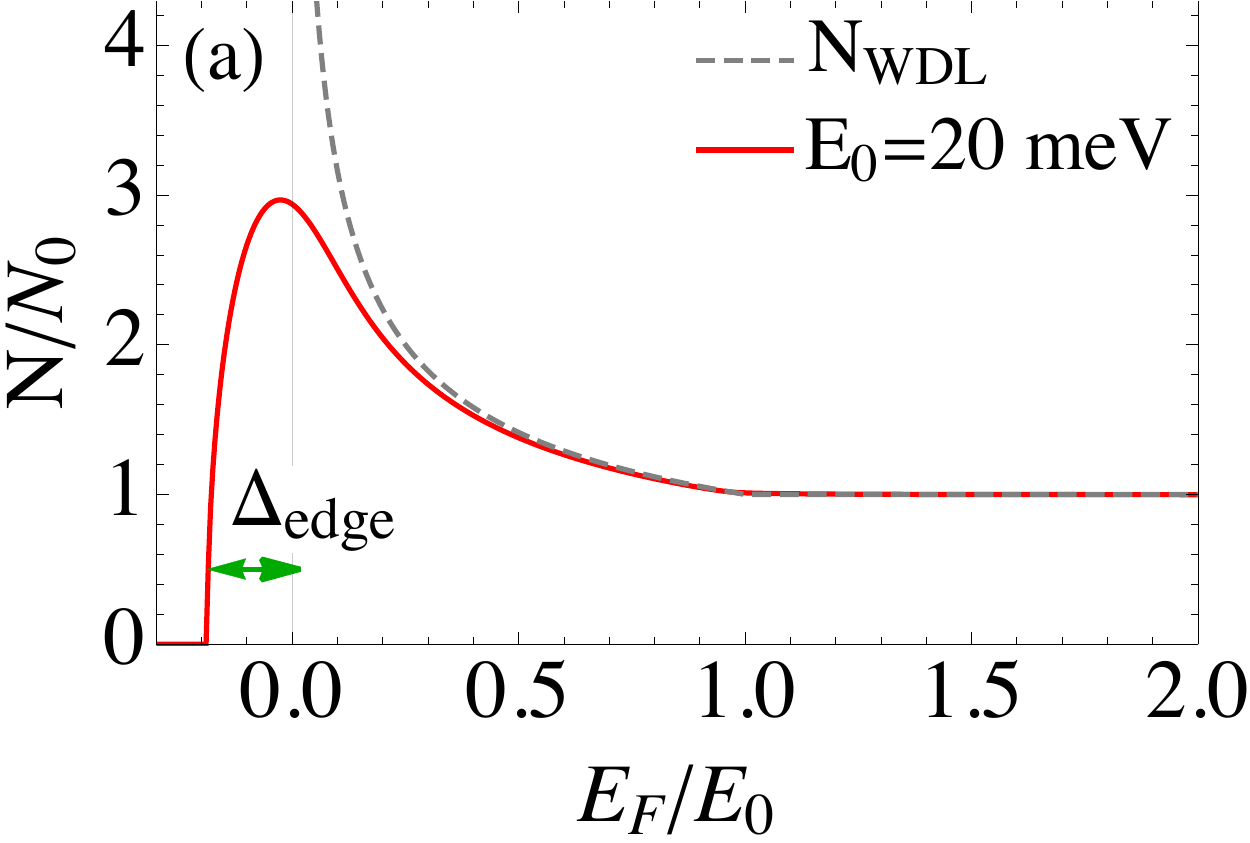}
\includegraphics[width=8cm]{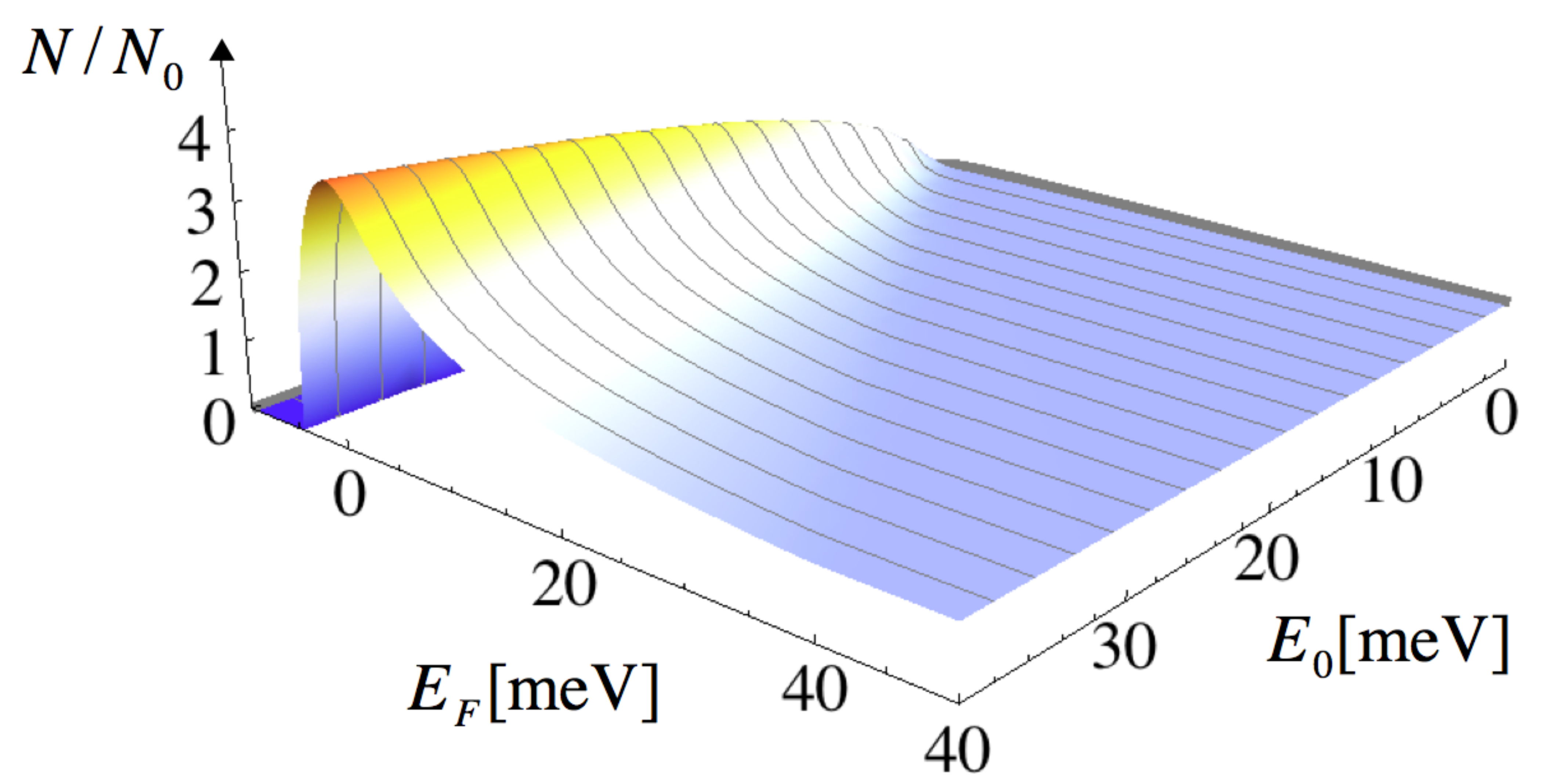}
\end{flushleft}
  \caption{DOS normalized to $N_0=m/\pi$ as a function of the bare Fermi energy for $E_0=40 \Gamma_0$ , $\Gamma_0=0.5meV$ and $m=0.7 m_e$. Comparison with the WDL results, see Table 1. 
  (b) Behavior of the DOS as a function of $E_F$ and $E_0$. Other parameters as in panel (a).  } 
 \label{smearing}
\end{figure}

\noindent A strong Rashba spin-orbit coupling has remarkable effects both on the single-particle and on the transport properties of a two-dimensional electron-gas.

\begin{figure}[t!]
\begin{center}
\includegraphics[width=0.4\textwidth]{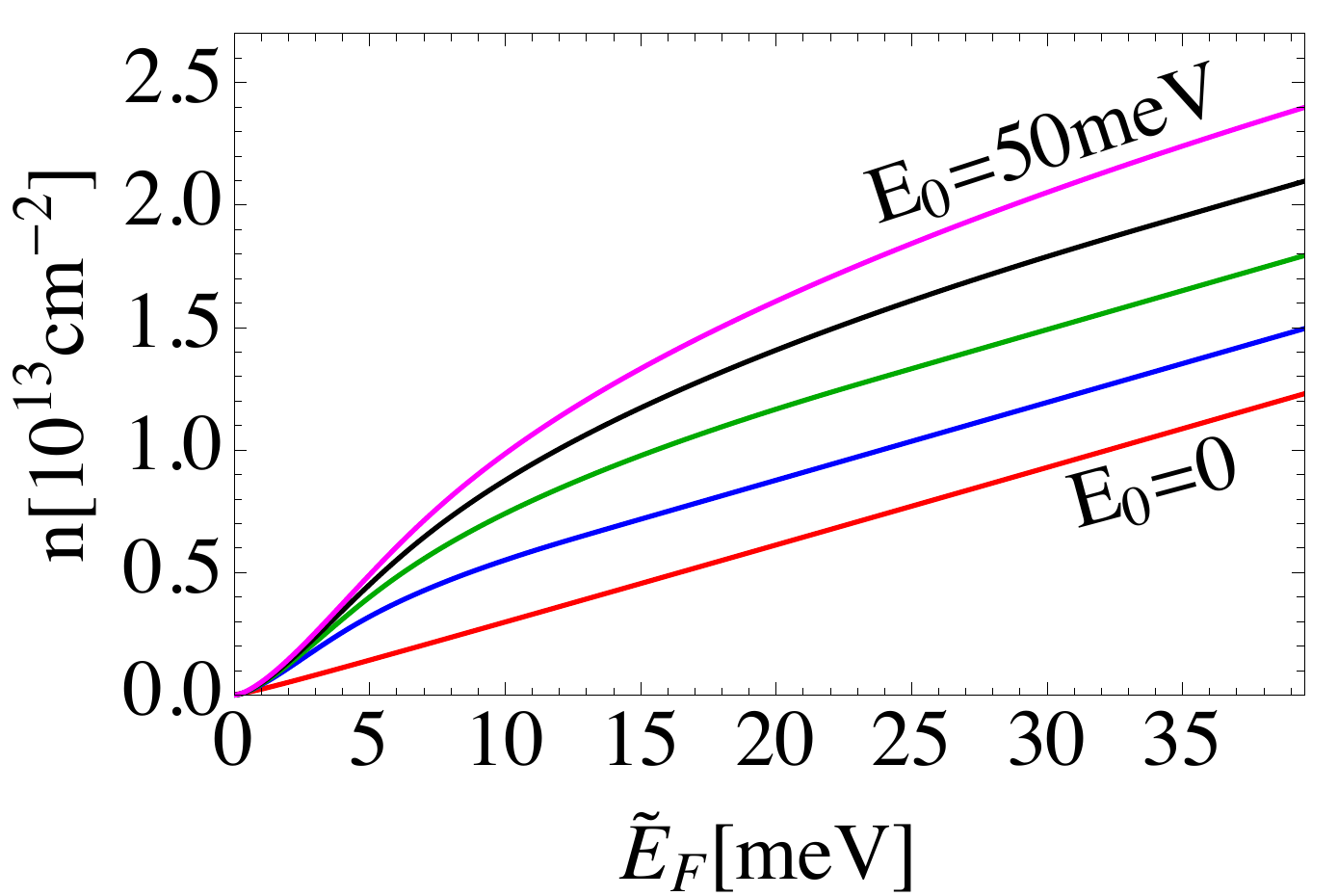}
\includegraphics[width=0.4\textwidth]{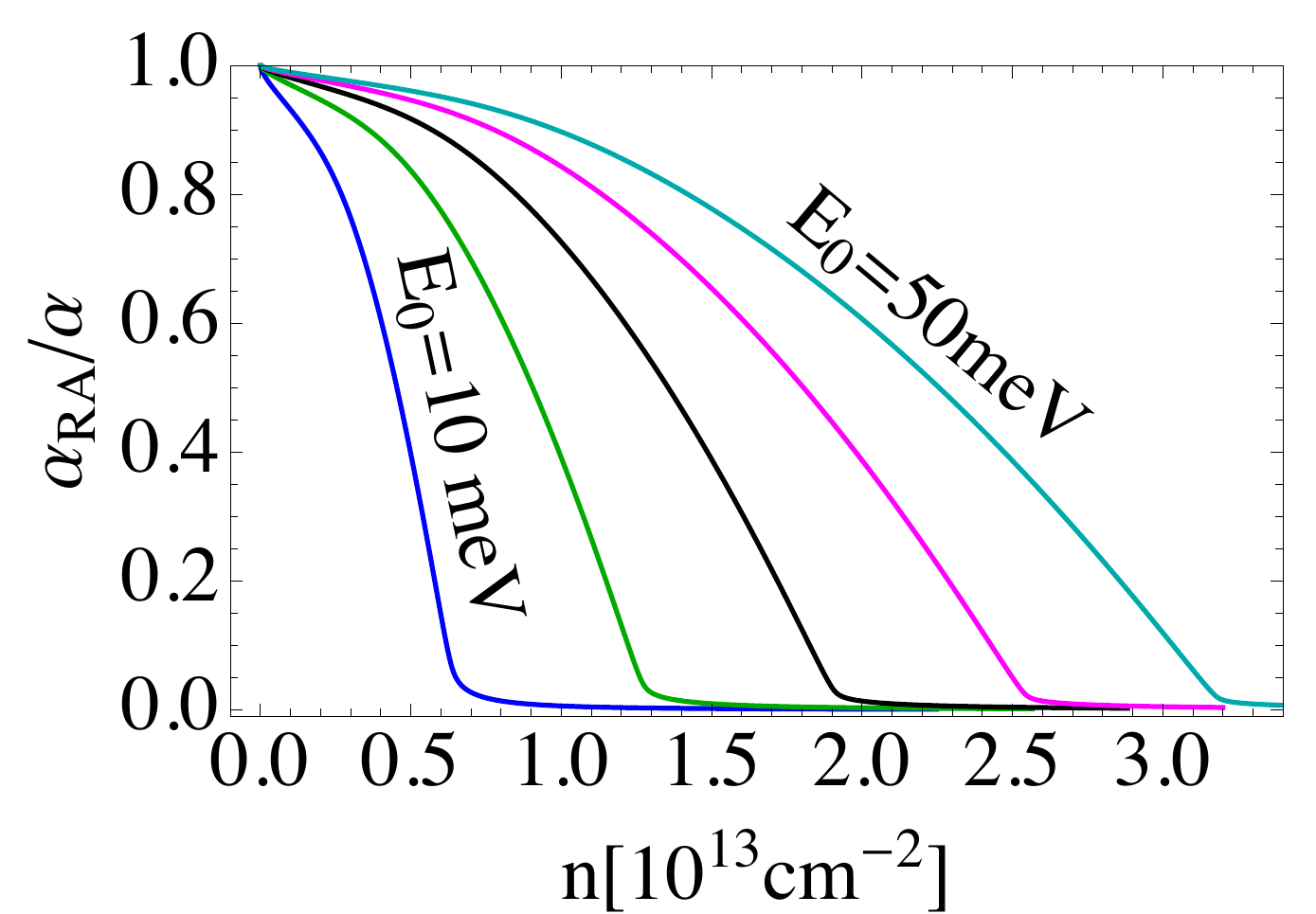}
\end{center}
  \caption{Dependence of the electronic density on the Fermi energy for different values of $E_0$ , $\Gamma_0=0.5 {\rm meV}$ and $m=0.7 m_e$. 
  (b) Behavior of the renormalized vertex as a function of  the density for different values of $E_0$. Other parameters as in panel (a).  } 
 \label{density}
\end{figure}

Let us first consider the effects on the single-particle properties. The vanishing of the quasi-particle velocity, $\vec{\rm v}_{{\bf p}s}$ at $|{\bf p}|=p_0$  yields a van-Hove singularity in the density of states $N(E)$ and in the quasi-particle scattering rate. As explained above, these two quantities are indeed related and,  as shown in Table 1, they  have both a square-root divergence as $E\rightarrow 0$.
The self-consistent DOS shows the signatures of the van Hove singularity, that is however smeared in the the presence of disorder,
as one can see in Figure \ref{smearing}(a), where we compare the approximate analytic DOS and the self-consistent one.
There we can also notice that a consequence of the smearing of the van Hove singularity is the presence of  a shift, $\Delta_{\rm edge}$, of the lower  band edge, so that  the lower ``effective'' band edge where $n=0$ is identified by
$\tilde E_F=E_F+\Delta_{\rm edge}=0$.   
The overall behavior of the density of states as a function of $E_0$ and $E_F$ is instead shown in Fig.\ref{smearing}(b).
It is interesting to note that, as first shown in Ref.\cite{cappelluti2007}, the change in the behavior of the density of states can be interpreted as an effective dimensional transition associated to the change of topology of the Fermi surface.
The transition is also accompanied by a change in the dependence of the total electronic density on $E_F$, shown in Figure \ref{density}(a).

We now come to the discussion of dc transport.
In Figure \ref{density}(b) we show the renormalized vertex $\tilde\alpha^{RA}$ as a function of the density. As one can see, while in the HD regime  we have $\tilde\alpha^{RA}=0$,  in the DSO regime the vertex becomes non-vanishing and it tends to $\alpha$ as the density goes to zero as summarized in Table 1.
The behavior of the renormalized vertex has been studied by various authors \cite{raimondi2005,Grimaldi2006,raimondi2001,schwab2002,raimondi2011} in the context of  the spin-Hall effect. In the stationary limit, in both regimes the spin-Hall conductivity of the Rashba model has to vanish \cite{dimitrova2005}. In the HD regime the vanishing of the renormalized vertex directly yields a vanishing  spin-Hall conductivity. In the DSO regime  instead  the spin-Hall conductivity vanishes due to a cancellation between on-shell RA and off-shell RR  contributions as first noted in Ref.\cite{Grimaldi2006}. 
The consequences of a  non-zero anomalous vertex are thus not visible in the  spin-Hall effect.
On the contrary, as we now show and discussed in Refs.\cite{brosco2016,brosco2017}, the presence of a non-zero anomalous vertex has important consequences on charge transport.
%
\begin{figure}[h!]
\begin{center}
\includegraphics[width=0.8\textwidth]{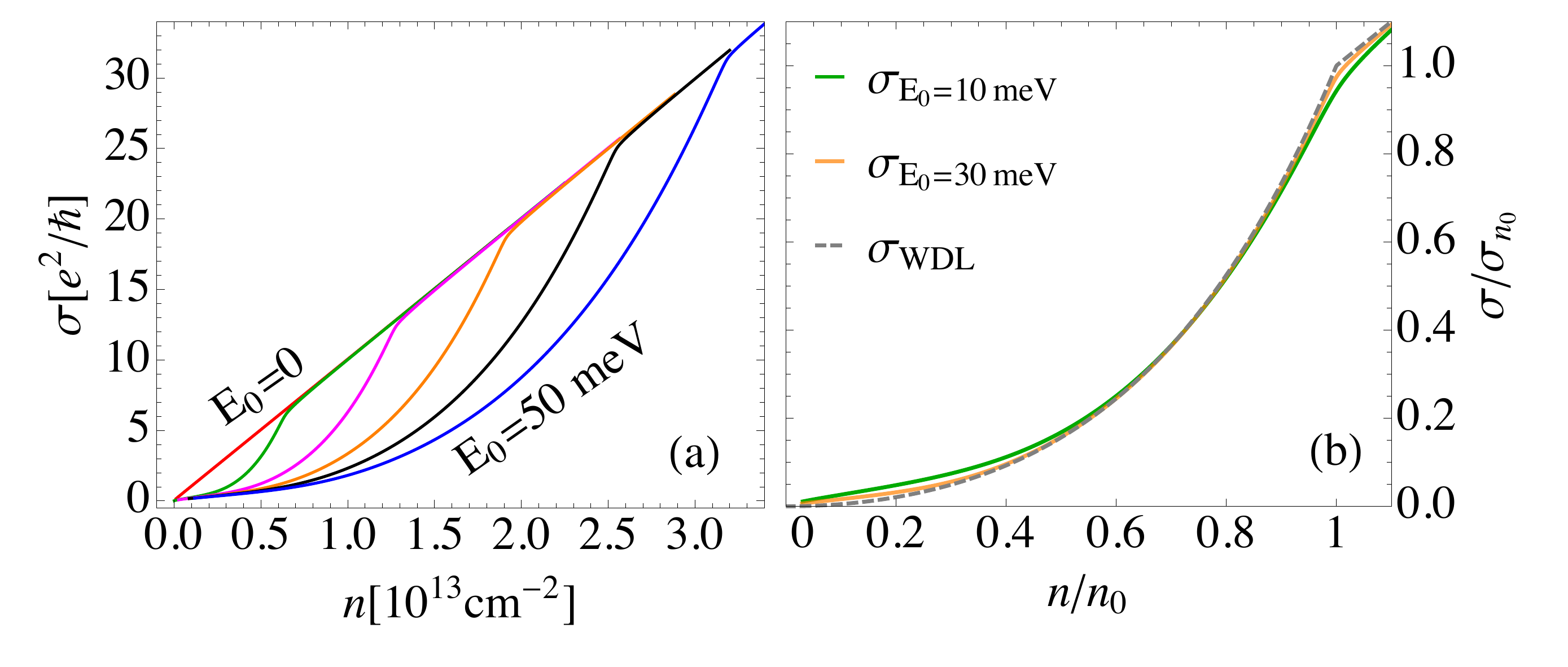}
\includegraphics[width=0.8\textwidth]{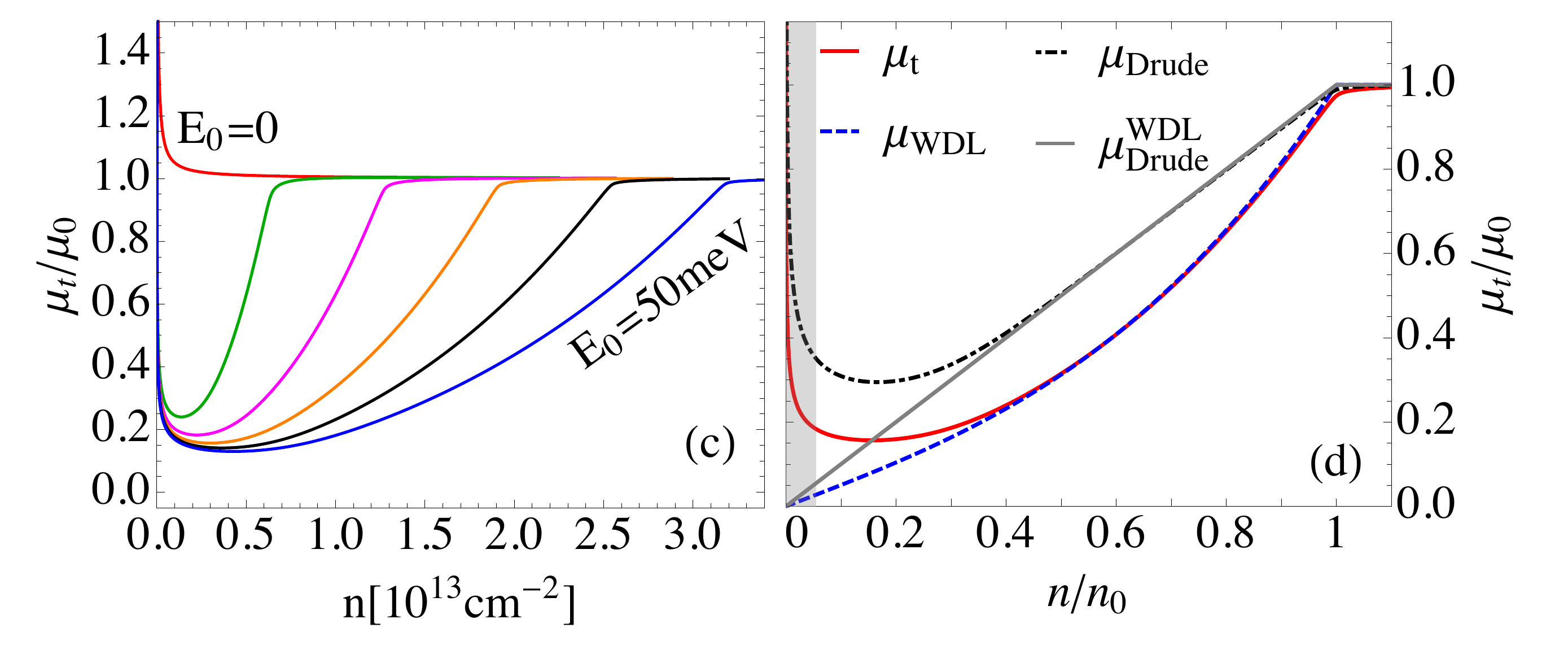}
\end{center}
  \caption{(a) Conductivity as a function of the density for different values of $E_0$. (b) Universal scaling of the conductivity as a function of $n/n_0$. (c) Mobility as a function of the density for different values of $E_0$. (d) Comparison between different theories for the mobility $E_0=30$meV. In all panels $\Gamma_0=0.5$meV and $m=0.7\, m_e$ } 
 \label{transport}
\end{figure}
%
The behavior of the conductivity as a function of the density for different values of $E_0$ is shown in Figure \ref{transport}(a). We notice that, while in the HD regime, the dc conductivity is simply given by Drude formula independently on how large is  $E_0$, in the DSO regime, instead, it is strongly suppressed.
This result has a two-fold origin, (i) the increase  of  the quasi-particle scattering rate; (ii) the  presence of  a non-zero renormalization of the anomalous velocity.
 
To understand this statement it is useful to analyze in more details Eqs.(\ref{pxxivel}) that defines the conductivity through equation (\ref{sigmaxx}). 
The $RR$ contribution becomes relevant only close to the band-edge while the behavior of the conductivity at finite densities is determined by $P^{RA}$. The latter is given by
\be\label{pxxRA}
P^{RA}_{xx}=P^{RA}_0+\frac{m\lf(1-A^{RA}\rg)}{\Gamma_0} \tilde \alpha^{RA}\left(\tilde \alpha^{RA}-\alpha\right)
\ee
where as in the previous section the superscript $RA$ means that the first argument is retarded while the second is advanced.
Starting from the above equation it is not difficult to understand why the conductivity is smaller than Drude result. Indeed one can show (see e.g.  the Appendix of Ref.\cite{brosco2017}) that 
 $P^{RA}_0$ always yields a Drude-like contribution to the conductivity, while the second term  is negative since $\tilde \alpha^{RA}\leq\alpha$ and all other terms are positive, in particular $1- A^{RA}>0$.

 As shown in Fig.\ref{transport}a as soon as the system enters the dominant SO coupling regime   the conductivity as a function of the density deviates from Drude law and it becomes sublinear, the transition happening at $n=n_0$. This behavior is well reproduced by the approximate  formula shown in Table 1, which yields a ``universal''  dependence of $\sigma/\sigma_{n_0}$ as a function of $n/n_0$,  as shown in Fig.\ref{transport}b. 
 The accuracy  of the analytical formula and the deviations from Drude law emerge more clearly if we consider the mobility  $\mu_t\equiv\sigma/(e\hbar n)$ plotted in Fig. \ref{transport}c.
 There we compare different theories for the mobility:
 \begin{itemize}
 \item Drude theory with the fully self-consistent quasi-particle scattering rate, {\sl i.e.} $\mu_{\rm Drude}\sim \tau/m$ with $\tau=1/(2\Gamma)$;
 \item Drude theory with the quasi-particle scattering rate estimated with WDL approximation  {\sl i.e.} $\mu_{\rm Drude}^{WDL}\sim \tau_{\rm WDL}/m$ with $\tau=1/(2\Gamma_{WDL})$
 \item the fully self-consistent theory, $\mu_t$;
 \item the WDL approximation to the mobility including vertex corrections,  $\mu_{WDL}=\sigma_{WDL}/n$.
 \end{itemize}
 We remark that WDL results are those summarized in Table \ref{table2}.
As one can see,  even accounting for the enhancement of the scattering rate at $E_F<E_0$, Drude formula, $\mu_{\rm Drude}$, fails both qualitatively and quantitatively to describe the behavior of the mobility in the DSO regime. Indeed using the results given in Table1, we obtain 
$$\mu_{\rm Drude}^{WDL}\sim \frac{\tau_0}{m}\cdot \frac{n}{n_0} \quad {\rm and}\quad \mu^{WDL}\sim\mu_{\rm Drude}^{WDL}\,\frac{1}{2}\lf(1+\frac{n^2}{n_0^2}\rg)$$
The additional suppression coming from the factor in parenthesis, due to vertex corrections, changes the qualitative behavior of the mobility and, as shown in Fig. \ref{transport}(d) is needed to accurately describe $\mu_t$ over a broad range of SO coupling strengths and densities. The deviations of  $\mu_{WDL}$  from $\mu_t$  at low density are mostly due to the wrong estimate of the elastic scattering rate given by the WDL. The divergence of the mobility in the extreme diluted limit (shaded region in in Fig.\ref{transport}(d))  is related to the vanishing of $\Gamma$ within first order self-consistent Born approximation.

To get further insight into the physical meaning of the different approximations, we now rewrite the RA response function in the chiral basis as the sum of an intra-  plus an inter-band contribution. To this end we go back to  Eq.(\ref{pxxivel}) that defines the structure of $P_{xx}^{RA}$ and we rewrite explicitly all the different terms as follows:
\bea
P^{RA}_{xx}=\frac{1}{2\cal V}\sum_{\bf p}\lf\{\lf[\lf(\frac{p^2}{m^2}+\a\tilde\a^{RA}\rg)(g_+^Rg_+^A+g_-^Rg_-^A)\rg]+\rg.\nn\\ \lf.\frac{p(\a+\tilde\a)}{m}(g_+^Rg_+^A-g_-^Rg_-^A)+\a\tilde\a^{RA}(g_+^Rg_-^A+g_-^Rg_+^A)\rg\}\nn.
\eea
Starting from the above expression it is straightforward to identify intra- and inter-band contributions to $P_{xx}^{RA}$ and recast it as follows:
$
P_{xx}^{RA}= P_{\rm intra}^{RA}+P_{\rm inter}^{RA},
$
 where
\bea 
P_{\rm intra}^{RA}&=&\frac{e^2}{2{\cal V}} \sum\nolimits_{\bf p s} \vec v_{{\bf p},s}\cdot  \vec V^{RA}_{{\bf p},s} g^R_s(p,0)g^A_s(p,0),\label{ARMV} \\
 P_{\rm inter}^{RA}&=&\frac{e^2}{2{\cal V}} \alpha\, \tilde \alpha^{RA}\sum\nolimits_{{\bf p}\, s\neq s'}g^R_s(p,0)g^A_{s'}(p,0).\label{BRMV}
\eea
The interband term, $P_{\rm inter}^{RA}$, is always negligible as long as  the disorder-induced broadening of the spectral functions does not overcome the energy difference between the chiral bands. We can then  focus on the intraband term only, so that in the weak disorder limit 
$\sigma_{dc}$ can be recast as: 
\be
\label{app}
\sigma_{dc}\simeq\frac{e^2}{4{\cal V}\Gamma}  \sum\nolimits_{\bf p s} \vec v_{{\bf p},s}\cdot  \vec V^{RA}_{{\bf p},s} \delta(E_p^s-E_F),
\ee
where we used also $g^R_s(p,0)g^A_s(p,0)=(\pi/\Gamma){\cal A}_s(p)\simeq (\pi/\Gamma)  \delta(E_p^s-E_F)$.
With some simple algebra, from the above equation is possible to recover the WDL expression for the conductivity given in Table 1. In addition, we notice that from Eq.(\ref{app})  the conductivity can be recast as a sum of two contributions coming from the two branches of the Fermi surface that, as explained in Section \ref{sect-tr-hel} correspond to different values of the transport helicity. 
In this way it is not difficult to show (see the Supplementary Material of  \cite{brosco2016} for more details)   that the total conductivity can be written in both DSP and HD regimes as:
\be
\sigma_{dc}\simeq\sum_{\eta}\sigma_\eta
\ee
with $\sigma_{\eta}=v_F\tau^{\rm tr}_{\eta}p_\eta/(4\pi)$ where $v_F=\sqrt{2E_F/m}$ while $p_{\eta}$ and $\tau^{\rm tr}_{\eta}$ denote respectively the Femi momentum and  and the transport scattering time in each transport helicity band.
As one can easily show the Fermi momentum is given by $p_{\eta}=|mv_F-p_0|$ while $\tau^{\rm tr}_\eta$  is defined as
\be
\tau^{\rm tr}_\eta=\tau \frac{V^{RA}_{\bf p\eta}}{v_{\bf p\eta}}=\tau p_{\eta}/\bar p_F
\ee
with $\bar p_F=mv_F$ for $E_F>E_0$ and $\bar p_F=mv_F$ for $E_F<E_0$. 
It is interesting to note that, as one may expect, the above expressions  and all the WDL results can also be derived using the Boltzmann approach without making use of diagrammatic perturbation theory, {\sl i.e.} Boltzmann approximation can be recovered from  the weak disorder limit of the quantum Kubo results, see {\sl e.g.} \cite{brosco2016}.

%
%
 

%

\section{Thermal signatures of the DSO regime}
The purpose of this Section is two-fold. On one hand we aim at clarifying how robust is unconventional transport  with respect to temperature-induced broadening. On the other hand we would like to  investigate at which extent signatures of Rashba spin-orbit coupling may be found also looking at the temperature dependence of transport and equilibrium properties.

Let us remark that, as discussed in a number of works, (see e.g. Refs.\cite{dassarma2015,gold1986})  to fully describe low-temperature charge transport in two-dimensional semiconductors one should include the effect of screened Coulomb interactions  along with the temperature dependence of the electronic polarizability. 
Here, however, we limit ourselves to  consider only non-interacting electrons and  short-range impurities,  we thus neglect screening effects that will be the subject of a future work.
Furthermore we neglect the effects of electron-phonon coupling.

\begin{figure}[t!]
\begin{center}
\includegraphics[width=12cm]{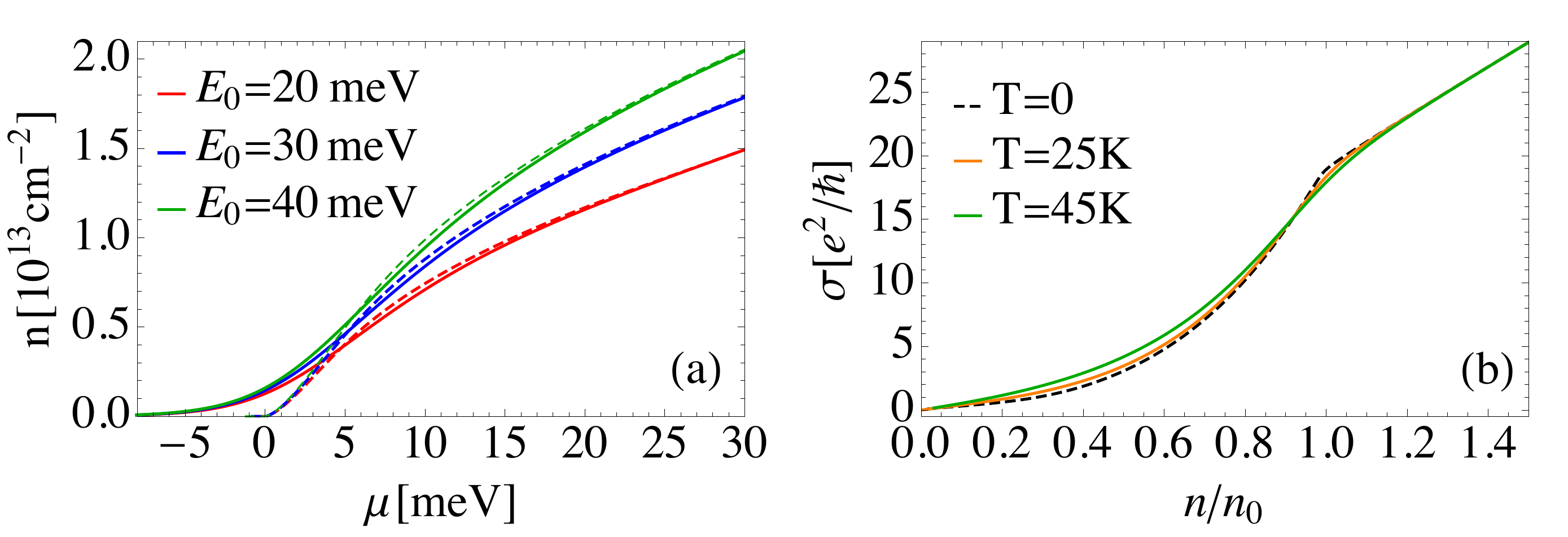}
\end{center}
  \caption{(a) Density as a function of the chemical potential at $T=0$ (dashed lines) and $T=30 K$ (solid lines) for different value of $E_0$.  (b) Conductivity as a function of the density for $E_0=30$ meV and different temperatures. Other parameters as in Fig.\ref{transport}.  } 
 \label{fig:muT}
\end{figure}

\begin{figure}[t!]
\begin{center}
\includegraphics[width=12cm]{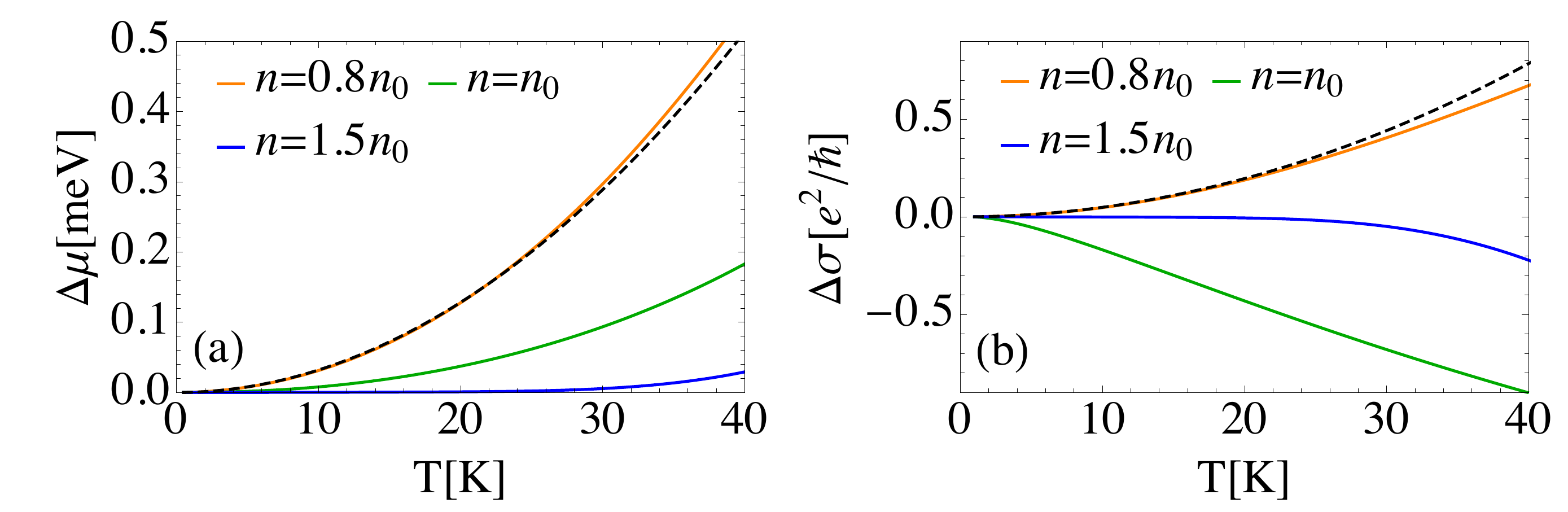}
\end{center}
  \caption{(a)Variation of the chemical potential, $\Delta\mu=\mu(T)-E_F$ as a function of the temperature for different fillings. (b) Variation of the conductivity, $\Delta\sigma=\sigma(T)-\sigma(T=0)$  as a function  of the temperature for different fillings. In both panels, $E_0=30$meV and  the dashed lines represent the Sommerfeld approximation given by Eq. (\ref{eq:muT2}) and Eq.\ (\ref{eq:sigmaT1}), respectively.} 
 \label{fig:musigmavsT}
\end{figure}
We start by looking at the smearing of the dependence of the density on the chemical potential, $\mu$, shown in Fig. \ref{fig:muT}(a). 
 The finite temperature results, represented by the solid lines, are obtained by integrating numerically the self-consistent density of states, as prescribed by Eq.(\ref{eq:nT}). To this end we evaluate the density of states on a wide range of energies, $W\gg E_0 \gg k_{\rm B} T$, with a suitably fine mesh $\delta\ve\ll k_{\rm B} T$. Indeed, as we shall discuss below, the behavior at finite temperature depends on the way the temperature $T$ induces the crossover from the DSO to the HD regime, so that some additional care is required to properly describe the competition between the three relevant energy scales set by $T$, $E_F$ and $E_0$. 
In Fig. \ref{fig:muT}(a) we notice that, as one may expect, finite temperature effects are most relevant close to the band-edge  and they remain significant in the DSO regime while they become negligible in the HD regime.
This can be understood  considering that, as explained above, due to the change in the Fermi surface topology, the transition from  the HD  to the DSO regime can be considered as a transition from an effectively two-dimensional  to a one-dimensional electron gas. As a result, similarly to what happen in standard 2DEGs, in the HD regime, even in the presence of a finite Rashba coupling, the effect of the temperature on the density is exponentially small.  On the contrary, in the DSO regime it remains finite analogously to what happens in one-dimensional electron gases.

Also the conductivity, whose finite temperature behavior can be obtained by  evaluating numerically Eq.(\ref{sigmaxx}) with the fully self-consistent $RR$ and $RA$ response functions, displays  a similar dichotomic behavior. Indeed as shown in Fig. \ref{fig:muT}(b), finite temperature effects are visible in the DSO regime while they become negligible in the HD regime.

In Fig.\ref{fig:musigmavsT}(a) and \ref{fig:musigmavsT}(b) we investigate the temperature dependence of the chemical potential and of the conductivity for different doping regimes. Specifically, in Fig.\ref{fig:musigmavsT}(a), we plot the difference  $\Delta\mu(T)=\mu(T)-E_F$, for three different fillings, namely, $n=0.8\, n_0$ and  $n=1.5\, n_0$, corresponding to the DSO and the HD regimes, and  $n=n_0$,  corresponding to the degeneracy point where the two bands touch. The results are represented by the solid lines and they are obtained by inverting numerically Eq.(\ref{eq:nT}) with the fully self-consistent density of states keeping the density fixed across the whole temperature range.
 Similarly,  in Fig.\ref{fig:musigmavsT}(b) the solid lines represent $\Delta\sigma=\sigma(T)-\sigma(T=0)$  and they are obtained by evaluating numerically the integral in Eq.(\ref{sigmaxx})  with the fully self-consistent response functions and  the chemical potential given by the inversion of Eq.(\ref{eq:nT}).
In the HD regime we notice that, as expected,  both the chemical potential and the conductivity show a rather weak dependence on the temperature. Thermal effects become instead much stronger at the degeneracy point and in the DSO regime, in particular for the conductivity.

\begin{figure}[h!]
\begin{center}
\includegraphics[width=12cm]{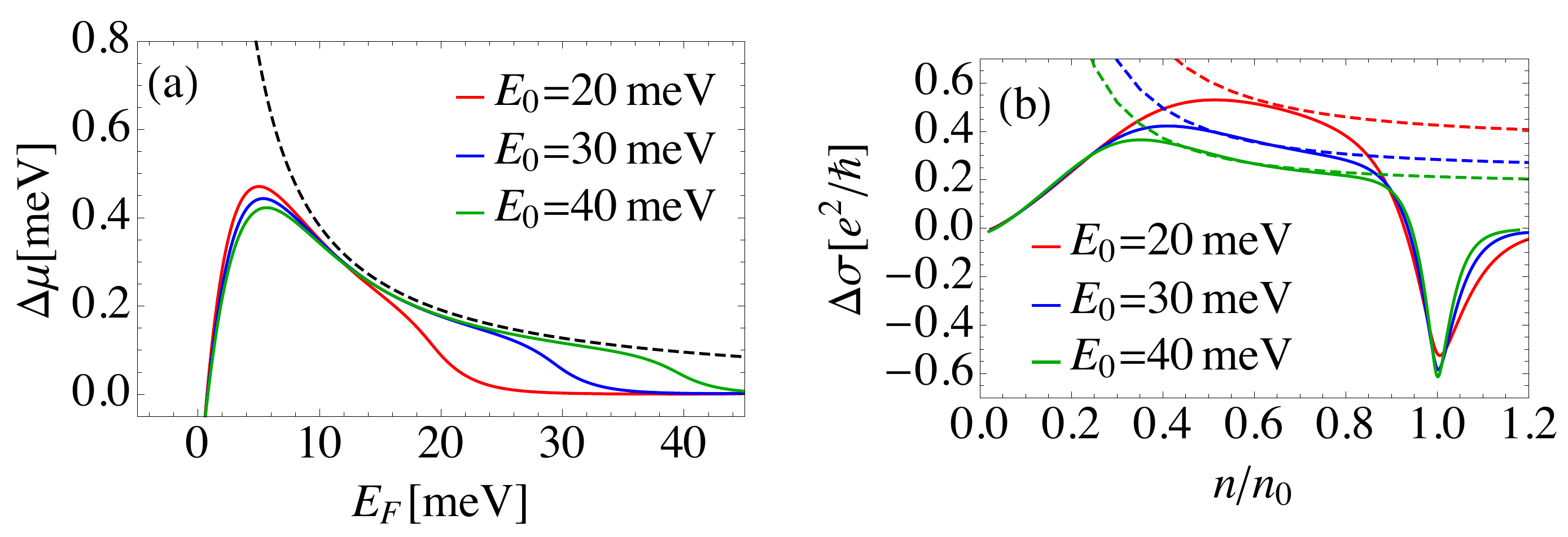}
\end{center}
  \caption{(a)Variation of the chemical potential, $\Delta\mu=\mu(T)-E_F$ as a function of the Fermi energy for different values of $E_0$. (b) Variation of the conductivity, $\Delta\sigma=\sigma(T)-\sigma(T=0)$ as a function of the density for values of $E_0$. In both panels the dashed lines represent Sommerfeld approximation.} 
 \label{fig:musigmaEFT}
\end{figure}

To gain some analytical understanding of the above results, we can estimate the integrals in Eqs.(\ref{eq:nT}) and (\ref{sigmaxx}) using Sommerfeld expansion \cite{ashcroft}.
By doing so it is not difficult to show that the chemical potential has the following approximate dependence on the Fermi energy and on the temperature
in the DSO regime:
\be\label{eq:muT}
\mu=E_F-\frac{N^\prime(E_F)}{N(E_F)}\frac{(\pi k_B T)^2}{6} \quad E_F<E_0
\ee
which, using the approximate expression for the density of states give in Table 1, yields
\be\label{eq:muT2}
\mu=E_F+\frac{(\pi k_B T)^2}{12 E_F}  \quad E_F<E_0
\ee
exactly the same result one would obtain in ordinary one-dimensional electron gases.
In Figure \ref{fig:musigmavsT}(a),  Sommerfeld approximation, {\sl i.e.} the r.h.s. of Eq.(\ref{eq:muT2}), is represented by the dashed black line. 
There we see that Sommerfeld approximation is generally very accurate and it fails only due to the finite energy range of the DSO regime, {i.e.}  very close to the band edge and close to the transition to the HD regime for $(E_0-E_F)/(k_BT)\sim 1$. The latter condition is  realized in Fig. \ref{fig:musigmavsT}(a) at $T\sim 30 K$. \\
Interestingly, Eq.(\ref{eq:muT2}) shows that thermal effects are universal, {\sl i.e.} they do not depend on the value of $E_0$ as long as the system is in the DSO regime. 
This fact also emerges quite clearly in Fig.\ref{fig:musigmaEFT}(a) where we plot the difference, $\Delta\mu$ between the chemical potential at $T=25 K$ and the Fermi energy as a function of the Fermi energy. We indeed see that  the approximation given by Sommerfeld expansion is accurate independently on the value of $E_0$ as long as the system is in the DSO regime. In this Figure we can also appreciate the ``dimensional nature'' of the transition between the DSO and the HD regime.

%
%
%
%
%

Let us now apply Sommerfeld expansion to describe the temperature  dependence of the conductivity.
To this end we start from Eq.(\ref{sigmaxx}), we integrate it by parts and we employ Sommerfeld expansion to estimate the integral. We recall that in this case the final result stems  from two factors: the smearing of the Fermi function and the temperature dependence of the chemical potential.
Including both effects we arrive at the following expression:
\be{\label{eq:sigmaT}}
\tilde \sigma_{\rm dc}(T)\sim\sigma_{\rm dc}(T=0)+\frac{(\pi k_B T)^2}{6}\lf[\sigma_{\rm dc}^{\prime\prime}(E_F)-\sigma_{\rm dc}^{\prime}(E_F)\frac{N^\prime(E_F)}{N(E_F)}\rg]
\ee
where we used Eq.(\ref{eq:muT}) for the chemical potential and  $\tilde \sigma_{\rm dc}(T)$ indicates the temperature dependent conductivity as given by Eq. (\ref{sigmaxx}) while $\sigma_{\rm dc}^{\prime\prime}$ and $\sigma_{\rm dc}^{\prime}$ are the derivative of the zero temperature conductivity with respect to $E_F$.
Using the expressions given in Table 1 and indicating as  $\Delta\sigma$ the difference between the finite temperature and the zero temperature conductivity, {\sl i.e.} $\Delta\sigma=\tilde \sigma(T)-\sigma(T=0)$,
we can recast Eq.(\ref{eq:sigmaT}) as follows:
\be{\label{eq:sigmaT1}}
\Delta\sigma\sim\frac{\pi (k_B T)^2}{24\Gamma_0 }\frac{6E_F+E_0}{ E_0E_F} \quad E_F<E_0
\ee
which in terms of the density becomes
\be{\label{eq:sigmaT2}}
\Delta\sigma\sim\frac{\pi (k_B T)^2}{4\Gamma_0 E_0 }\lf(1+\frac{n_0^2}{6n^2}\rg) \quad E_F<E_0.
\ee

In spite of the many approximations done in its derivation, Eq. (\ref{eq:sigmaT1}) provides a rather accurate description of the temperature dependence of the conductivity in the DSO regime. This can be seen in particular in Fig. \ref{fig:musigmavsT}(b) where the analytical results represented by the dashed black line follow quite closely the numerical ones.
A more comprehensive view can be instead gained by looking at Fig. \ref{fig:musigmaEFT}(b).   There we compare the thermal average of the self-consitent conductivity, given by the numerical evaluation of Eq.(\ref{sigmaxx}), and the approximation provided by Sommerfeld expansion, (solid and dashed lines respectively) across a wide range of densities.
We notice that, as expected, Sommerfeld approximation is accurate only deep in the DSO regime, while it does not include band-edge effects as well as the transition from the DSO to the HD regime which gives rise to a negative correction to the conductivity with increasing temperature, as also evident from Fig.\ref{fig:muT}.

\section{Conclusions}
In summary, in the present manuscript we outlined the behavior of the cd transport properties of the Rashba metal in the presence of point-like impurities, which represent the dominant source of scattering in the low-temperature regime. As it was previously discussed in Ref.\ \cite{brosco2016}, already at $T=0$ strong Rashba interaction  is able to induce a remarkable deviation of the dc conductivity from the standard Drude-like result. Such deviations are observable in the so-called DSO regime, i.e. when the Rashba energy $E_0$ overcomes the fermi energy $E_F$. In this case $\sigma_{dc}$ shows a sub-linear dependence on the electron density, that could be in principle tested by direct measurement of the transport in gated devices, where the density can be tuned continuously. In the present manuscript we investigated how robust are these finding as a function of the temperature. In general, whenever $E_0$ remains the dominant energy scales the anomalous density dependence in the DSO regime is preserved, as shown explicitly in Fig.\ \ref{fig:muT}b. 
In addition, systems in the DSO regime can also show a sizable temperature such temperature dependence may provide an alternative method to estimate the value of $E_0$. Measuring such temperature dependence may be used to estimate the value of $n_0$  and thus $E_0$, once the density is known.




\end{document}